# Increased Coupling in the Saliency Network is the main cause/effect of Attention Deficit Hyperactivity Disorder

Xiaoxi Ji[1,#], Wei Cheng[1,#], Jie Zhang[1], Tian Ge[1,2],

Li Sun[3], Yufeng Wang[3], Jianfeng Feng[1,2,*]

[1]Centre for Computational Systems Biology, School of Mathematical Sciences, Fudan University, Shanghai 200433, PR China;

[2]Department of Computer Science, University of Warwick, Coventry CV4 7AL, UK.

[3]Institute of Mental Health, Peking University, Beijing 100191, China

[#]These authors contributed equally to this work.

*Correspondence should be addressed to

Jianfeng Feng, email: jianfeng64@gmail.com

# ABSTRACT

To uncover the underlying mechanisms of mental disorders such as attention deficit hyperactivity disorder (ADHD) for improving both early diagnosis and therapy, it is increasingly recognized that we need a better understanding of how the brain's functional connections are altered. A new brain wide association study (BWAS) has been developed and used to investigate functional connectivity changes in the brains of patients suffering from ADHD using resting state fMRI data. To reliably find out the most significantly altered functional connectivity links and associate them with ADHD, a meta-analysis on a cohort of ever reported largest population comprising 249 patients and 253 healthy controls is carried out. The greatest change in ADHD patients was the increased coupling of the saliency network involving the anterior cingulate gyrus and anterior insula. A voxel-based morphometry analysis was also carried out but this revealed no evidence in the ADHD patients for altered grey matter volumes in the regions showing altered functional connectivity. This is the first evidence for the involvement of the saliency network in ADHD and it suggests that this may reflect increased sensitivity over the integration of the incoming sensory information and his/her own thoughts and the network as a switch is bias towards to the central executive network.

# INTRODUCTION

At the brain circuit level, most of what we understand about ADHD and its biological abnormalities during the resting state comes from fMRI studies targeting changes in a small number of brain regions, as recently reviewed in (1, 2). These studies have suggested the involvement of a default mode network including the dorsal anterior cingulate cortex and some subcortical areas such as amygdala and putamen (3-7). However, the conclusions drawn from these studies are based on either seed-based analysis or independent component analysis (ICA) and are questionable in spite of the wide and successful application of such methods in the analysis of resting state fMRI data (8-13). Seed-based analysis is a hypothesis-driven approach which means the foci (seeds) of the disorder must be specified a priori. It is therefore a biased approach lacking a global and independent view (14). With the ICA approach it is assumed that the human brain is composed of independent components whereas in reality different parts of the human brain undoubtedly work in a coordinated fashion. Hence, given the complexity and multiple causes of ADHD together with variability between individuals, a novel, unbiased approach is urgently called for which identifies key pathway changes in a holistic manner.

In (15), we have adopted a third and holistic approach aiming to un-ambiguously identify the key connections which are modified in the brains of depression patients. However, the population is quite small which refrains us from using a proper association study since no functional connectivity can survive after a multi-comparison correction. Here we have further developed a new brain wide association study (BWAS) and applied it to by far the largest population with ADHD in the literature (see Fig. 1). Voxel-based morphometry analysis was also carried out to assess whether gray or white matter volume changes occurred in those brain regions of ADHD patients showing altered functional connectivity.

The most significant differences between ADHD patients and healthy populations occurred in inter-community rather than intra-community links (15). The greatest change was in the salience network, comprising the anterior insula and anterior cingular gyrus (16), (see Fig. 2), where the coupling between these regions was increased in ADHD patients in left-hemisphere. It has been proposed that the primary role of the salience network is the integration of sensations, internally generated thoughts and information about goals and plans in order to update expectations about the internal and external milieu and, if necessary, to allow action to be initiated or modified (17). The salience network serves as a switch between internal status (default mode network) and central executive network (17). Our finding here suggests that in ADHD, due to the increasing of the coupling in the salience network, a patient is more sensitive to external stimuli or internal thoughts and the switch could bias toward central executive network. This is the first time that changes in the functional connectivity of the salience network have been identified in the brains of ADHD patients, although its role has been reported in schizophrenia and many other cognitive tasks (18). Actually, our result is also the first one to demonstrate that the salience network does play a cardinal role in psychosis (18) .

# RESULTS

## Canonical template

The six-community structure constructed for the whole brain from 253 healthy subjects is shown in Fig. S1. Each link represents a significant link between two brain regions, with their full names listed in Table 1. For clarity, we do not specify the left and right for each region and the existence of the six communities in the whole brain is as reported before (15). We have observed this same structure in an even larger population of around 400 people (data from Cambridge USA and Beijing publicly available in (19), results not shown). The six communities correspond to six Resting State Networks (RSN) which can be identified in terms of broad functions and can be classified as a default mode network (DMN) (RSN1), an attention network (RSN2), a visual recognition network (RSN3), an auditory network (RSN4), sensory-motor areas (RSN5) and a subcortical network (RSN6).

**PU and NYU patients** For both PU and NYU patients, functional maps were constructed and compared with those for the healthy subject group. Comparing the patient network with the canonical template from healthy subjects is described in detail in (15). In order to rank the significance of the change for each link, a score is defined as follows for each particular link:

$$S = \frac{L_H}{N_H} - \frac{L_P}{N_P}$$

where $S$ is the score for a particular link, $L_P$ is the number of this link present in the individual networks of ADHD, $N_P$ is the total number of patients, $L_H$ is the number of this link present in the individual networks of normal controls, and $N_H$ is the total number of healthy controls. We carried out a BWAS on total scores for different circuits to assess significance of changes in ADHD patients. After applying BWAS approach to our network, there are 9 links as in Table 2 which are statistically significant after the meta-study approach by pooling the two groups together. It can be seen from these figures that the strongest evidence for enhanced connectivity compared with control subjects in both PU and NYU is that between the insula and anterior cingulate gyrus in left brain hemispheres ($S = -0.12$ with $p = 5.8 \times 10^{-6}$, see below for the calculation of $p$-value). Thus the link which between the two main components of the saliency network have become largely coupled in patients. Connections to the left ORBmid from the left IPL ($S = 0.12$ with $p = 1.2\times10^{-4}$) and those from ORBmed right from PCG left ($S = 0.09$ with $p = 2.2\times10^{-4}$), are also increased in control group together with links from the left TPOsup to the right TPOsup ($S = 0.11$ and $p = 2.7\times10^{-4}$). In additional to these four changed circuits, we have also found that increase of coupling in control group between REC (left) and ANG (left) with a score of 0.086 and $p = 3.4\times10^{-4}$. Increased coupling in patients group for other four circuits are observed: they are MFG (left) and ACG (right), ORBsup (left) and ORBmed (left), ITG (right) and PHG (right), and finally AMYG (right) and HIP (right), see Table 2 for more details.

In order to validate our findings, we have further carried out an (semi-independent) study for both PU group and NYU group, with a pre-fixed *p*=0.05. 176 links (without correction) are found to be significant in PU group and 102 links in NYU group. The common links among these three studies (meta-analysis, PU, NYU) are four links: they are the saliency network, the ORBmed right to the left PCG ,the left TPOsup to the right TPOsup and the right ITG and the right PHG. Different ways of applying BWAS approach to the two datasets have been used, as shown in supplemental materials and further confirmed our findings. An alternative way of truly independent replicate method is worked out as described in Supplemental materials (Fig. S2) and we further confirmed that the sali-ence network has increased its links in patients, although the other links can not survive after correction.

Voxel-based morphometry analysis revealed no significant ($p > 0.05$ t-test with Bonferroni correction) grey matter volume reductions in these pathways (See Supplementary Tables S1 for details of the grey/white matter volumes of the ROIs involved in these pathways for both patients and normal controls are listed).

## Source Locations

After locating the main changes between healthy controls and ADHD patients, we now move a step forward to identity the exact location of the changed coupling between anterior cingulate gyrus and insula. Actually, it is reported that the saliency network is mainly involved the anterior insula (18). To the end, for each pair of voxels in insula and anterior cingular gyrus, we calculate the correlation coefficients between healthy controls and patients. The remaining significant volumes are plotted in warm color as in Fig. 3. The centre coordinates for the left insula and the left anterior cingulate gyrus source are (-38, 24, 5), (-14, 48, 13) in MNI space respectively, which are denoted by green stars. As is clearly shown, the involved part of insula in enhanced coupling lies at the anterior part of the insula.

## Correlation between various scores

After locating the most significant changed voxels (regions), we now intend to link these features with various other factors (scores, illness durations, IQ etc.). We have tried various quantities and, interestingly, the correlations are always very weak, possibly indicating that current diagnosis methods are not very consistent or the contributions to ADHD are multi-factors. The partial correlation coefficient between any pair of ROIs was defined as the minimum partial correlation coefficient of these two regions conditioning on any third region. Further, Partial correlation analyzes were performed between the IQ scores and the minimum partial correlation coefficients, regressing out the age and gender factors. The link between TPO left to right, however, is significantly correlated with IQ of patient groups, with a correlation coefficient being 0.2015 and $p$=0.0016.

# METHODS

## Subjects

Our data was provided by the ADHD-200 Consortium (http://fcon_1000.projects.nitrc.org/indi/adhd200/) for the global competition to develop classification for ADHD diagnosis. We choose two districts, New York University Medical Center (NYU) and ourselves (Peking University, PU), both groups provided the largest samples. Consistent with the policies of the 1000 Functional Connectome Project, data usage is unrestricted for non-commercial research purposes.

We used 502 resting-state fMRI (158 females, 344 males; ages: 7-18 years old; mean age 11.61 ± 2.49 years), 253 of which were obtained from typically developing individuals (control) (114 females, 139 males; mean age 11.73 ± 2.50 years) and 249 in children and adolescents with ADHD (44 females, 205 males; mean age 11.49 ± 2.47 years) from the ADHD-200 datasets. Accompanying phenotypic information includes: diagnostic status, ADHD measure, age, sex, intelligence quotient (IQ) and so on.

**Peking University (PU)** 245 children participated in the experiment from Institute of Mental Health, Peking University and National Key Laboratory of Cognitive Neuroscience and Learning, Peking University, Beijing, China. They are 143 typically developing individuals (control) (59 females, 84 males; mean age 11.43 ± 1.86 years; mean IQ 118 ± 13.15 scores) and 102 patients with ADHD (10 females, 92 males; mean age 12.09 ± 2.04 years; mean IQ 106.03 ± 13.10 scores), including 38 ADHD-C, 1 ADHDHI and 63 ADHD-I. Further, for 102 patients with ADHD, 71medication-naive patients and other 31patients are not medication naive. All subjects scored intelligence quotient (IQ) on Wechsler Intelligence Scale for Chinese Children-Revised (WISCC-R) and scored ADHD index on 18-item version of ADHD Rating Scale IV (ADHD-RS) (mean index 50.38 ± 8.39 scores). Study participants with the diagnosis of ADHD were initially identified using the Computerized Diagnostic Inter-view Schedule IV (C-DIS-IV). Upon referral for participation to the study participation, all participants (ADHD and controls) were evaluated with the Schedule of Affective Disorders and Schizophrenia for Chil-dren—Present and Lifetime Version (KSADS-PL) with one parent for the establishment of the diagnosis for study inclusion. The ADHD Rating Scale (ADHD-RS) IV was employed to provide dimensional measures of ADHD symptoms. Additional inclusion criteria included: (i) right-handedness, (ii) no lifetime history of head trauma with loss of consciousness, (iii) no history of neurological disease and no diagnosis of either schizophrenia, affective disorder, pervasive development disorder, or substance abuse and (iv) full scale Wechsler Intelligence Scale for Chinese Children-Revised (WISCC-R) score of greater than 80. Psychostimulant medications were withheld at least 48 hours prior to scanning. All research was approved by the Research Ethics Review Board of Institute of Mental Health, Peking University. Informed consent was also obtained from the parent of each subject and all of the children agreed to participate in the study. The two scans were at least 2 days apart, and each scan was taken 1 hour after either 10mg MPH administration or placebo (Vitamin B6, 10mg). All the patients had not received stimulant treatment for at least two days before the first scan, and were asked not to take any stimulant between two scans. The control boys were scanned once without MPH or placebo taken for ethical reasons. Only placebo scans were used for the present study.

**NYU** 257 participants were recruited from Phyllis Green and Randolph Cowen Institute for Pediatric Neuroscience at the Child Study Center, New York University Langone Medical Center, New York and Nathan Kline Institute for Psychiatric Research, Orangeburg, New York, USA. 110 typically developing individuals (control) (55 females, 55 males; mean age 12.12 ± 3.12 years; mean IQ 111 ± 14.14 scores) and 147 patients with ADHD (34 females, 113 males; mean age 11.07 ± 2.66 years; mean IQ 105.82 ± 14.09 scores), including 95 ADHD-C, 2 ADHD-HI and 50 ADHDI. Further, 31 patients are medication-naive, 28 patients are not medication naive, and medication status of other 88 patients are not clear. All subjects scored intelligence quotient (IQ) on Wechsler Abbreviated Scale of Intelligence (WASI) and scored ADHD index on Conners' Parent Rating Scale-Revised, Long version (CPRS-LV)(mean index 71.86 ± 8.64 scores). Psychiatric diagnoses were based on evaluations with the Schedule of Affective Disorders and Schizophrenia for Children—Present and Lifetime Version (KSADS-PL) administered to parents and children and the Conners' Parent Rating Scale-Revised, Long version (CPRS-LV). Intelligence was evaluated with the Wechsler Abbreviated Scale of Intelligence (WASI). Inclusion in the ADHD group required a diagnosis of ADHD based on parent and child responses to the KSADS-PL as well as on a T-score greater than or equal to 65 on at least one ADHD related index of the CPRS-R: LV. Psychostimulant drugs were withheld at least 24 hours before scanning. Inclusion criteria for TDC required absence of any Axis-I psychiatric diagnoses per parent and child KSADS-PL interview, as well as T-scores below 60 for all the CPRS-R: LV ADHD summary scales. Estimates of FSIQ above 80, right-handedness and absence of other chronic medical conditions were required for all children. The property of the datasets is summarized as in Table 3.

## Imaging acquisitions and data preprocessing

All functional imaging data were acquired using AFNI (an acronym for Analysis of Functional NeuroImages) and FSL (http://www.fmrib.ox.ac. uk/ fsl/) running on the Athena computer cluster at Virginia Tech's ARC (http://www.arc.vt.edu/) by Cameron Craddock. The Athena pipeline is primarily focused on resting state fMRI processing but also preprocesses the T1 images in order to achieve a high quality transformation between subject space and MNI space. Before functional images preprocessing, the first 4 volumes were discarded (AFNI) to allow for scanner. Briefly, the remaining functional scans were first corrected for within-scan acquisition time differences between slices, and then realigned to the middle volume to correct for interscan head motions. Subsequently, preprocessed resting state fMRI data, the functional scans were spatially normalized to a standard template (Montreal Neurological Institute) and resampled to 4

mm×4 mm×4mm voxel resolution. After normalization, the Blood Oxygenation Level Dependent (BOLD) signal of each voxel was first detrended to abandon linear trend and then passed through a bandpass filter (0.009 Hz < f < 0.08 Hz) to reduce low-frequency drift and high-frequency physiological noise. Finally, nuisance covariates including head motion parameters, global mean signals, white matter signals and cerebrospinal fluid signals were regressed out from the Blood Oxygenation Level Dependent signals. Time courses were extracted from filtered preprocessed resting state data using the automated anatomical labeling (AAL).

All voxel based morphometry were processed (grey matter and white matter) with DARTEL based on the SPM software package in Burner pipeline from the ADHD-200 consortium. The Burner pipeline includes modulated and normalized grey matter maps generated using SPM8 (University College London, UK; http://www.fil.ion.ucl.ac.uk/spm). There are three steps in Burner pipeline. Firstly, T1 images were segmented into grey matter and white matter probability maps using "New Segmentation" in SPM8. The tissue maps were rigidly aligned (translation and rotation). Next, Inter subject normalization was performed using DARTEL toolbox in SPM8. Images were iteratively registered to the group average (population template), and template was updated iteratively. This resulted finer and finer registration. Lastly, the registration pa-rameters were applied to each image to transform each image into the space of population average. Modulation was applied to conserve the global tissue volumes after normalization.

**PU** one resting-state fMRI scans. Participants were asked to relax, stay still, and either keep their eyes open or close. A black screen with a white fixation cross was displayed during the scan. One high-resolution T1-weighted mprage (magnetization prepared rapid acquisition gradient echo), defaced to protect patient confidentiality. All functional imaging data and T1-weighted images were acquired using SIEMENS TRIO 3-Tesla. A total of 232 volumes of echo planar images were obtained axially (Axial scanned; 30 slices; slice order: 2:2:30, 1:2:29; TR, 2000 ms; TE, 30 ms; slice thickness, 4.5 mm; slice spacing, 0 mm; flip angle, 90° ; FOV = 220×220; Matrix = 64×64, 8 minutes, 240 TR). Subjects were scanned using 1 of 5 different sagittal t1-mprage protocols. The parameters are below: (1). 192 slices, TR= 2000 ms, TE= 3.67 ms, Inversion Time= 1100 ms, slice thickness= 1 mm, Flip angle = 12°, FOV = 240×240, Matrix = 256×256; (2).128 slices, TR = 1950 ms, TE = 2.6 ms, Inversion Time = 900 ms, slice thickness= 1.3 mm, Flip angle = 10°, FOV = 240×256, Matrix = 240×256; (3).128 slices, TR = 2530 ms, TE = 3.37 ms, Inversion Time = 1100 ms, slice thickness = 1.33 mm, Flip angle = 7°, FOV = 256×256, Matrix = 256×256; (4).176 slices, TR = 1770 ms, TE=3.92 ms, Inversion Time = 1100 ms, slice thickness = 1 mm,

Flip angle = 12°, FOV = 256×256, Matrix = 512×512 ;(5).144 slices, TR = 845 ms, TE = 2.89 ms, Inversion Time = 600 ms, slice thickness =1.3 mm, Flip angle＝8°, FOV = 261×261, Matrix = 256×256.

**NYU** one or two resting-state fMRI scans. During acquisition, participants were asked simply to remain still, close their eyes, think of nothing systematically and not fall asleep. A black screen was presented to them. A total of 172 volumes of echo planar images were obtained axially (33 slices; TR, 2000 ms; TE, 15 ms; slice thickness, 4 mm; flip angle, 90°; FOV = 240×240; Matrix = 80×80, 6 minutes). One high-resolution T1-weighted mprage (magnetization prepared rapid acquisition gradient echo), defaced to protect patient confidentiality. The scan parameters are 128 slices, TR = 2530 ms, TE = 3.25 ms, Inversion Time = 1100 ms, slice thickness = 1.33 mm, Flip angle = 7°, FOV = 256×256, Matrix = 256×256. An automated anatomical labeling atlas was used to parcellate the brain into 90 regions of interest (ROIs) (45 in each hemisphere). The names of the ROIs and their corresponding abbreviations are listed in Table1.

## Construction of whole-brain functional network

After data preprocessing, the time series were extracted in each ROI by averaging the signals of all voxels within that region. Pearson correlation coefficients between all pairs of ROIs were first calculated. Significant correlations were detected with a $p$ value smaller than 0.01. A 90×90 correlation matrix was obtained for each subject. However, significant correlation between two ROIs may be spurious, i.e. a by-product of the correlations of the two ROIs with a third region. To find out whether the correlation for the two ROIs is genuine, the third ROI should be kept constant. Statistically, this problem can be tackled by means of a partial correlation test. In such a test, the effects of the third ROI upon the relation between the other two ROIs are eliminated. By calculating partial correlation coefficients between all pairs of ROIs with all the remaining ROIs being controlling variables, a $90 \times 90$ correlation matrix was obtained for each subject with a $p$ value smaller than 0.01.

## Community mining algorithm

A network community generally refers to a group of vertices within which the connecting links are dense but sparse in between. In this study, a community structure of the functional network of the brain corresponds to groups of brain regions that have similar functions and dense functional connectivity with each other. Our former developed community mining algorithm described in (20) tries to explore the notion of network modularity by means of understanding the dynamics of the network, which can naturally reflect the intrinsic properties of the network with modularity structure and exhibit local mixing behaviors. Based on large deviation theory (20), this algorithm sheds light on the fundamental significance of the network communities and the intrinsic relationships between the modularity and the characteristics of the network.

**Brain-wide Association Study**

BWAS approach is very similar to the GWAS approach. In GWAS, we associate genes with disease, with a p-value which should be significant enough to survive a correction (p in the scale of $5\times10^{-8}$). Here we intend to associate altered functional connectivity to brain disorders and hence a single link in BWAS is equivalent to a gene in GWAS. Depending on the number of ROIs, our *p*-value could change accordingly. In our case, we restrict ourselves to 90 ROIs and in the following to work out the p-value which could survive after correction.

Assume there is one group of subjects suffering from some mental disease, for example ADHD and another group of matched healthy controls. Denote these two groups as P and H respectively. For each subject, the whole brain is parcellated into *N*=90 regions and the representative signal for each region has been extracted in each ROI. The 90×90 correlation matrix as mentioned before was obtained for each subject. The population level network can be obtained by summarizing all individual networks in ADHD and Control groups respectively and thresholding them into binarized matrices. Now our task is to detect those links, or circuits, that appear at significant different frequencies in patients and healthy controls. Let's start from a particular link. Denote the total number of patients and healthy subjects as $N_P$ and $N_H$ respectively, and assume that this link appears at probability $p$ in patients, and $q$ in healthy controls. Further assume the independence among subjects. Then the numbers of patients/healthy controls with this link follow binomial distributions $B(N_P p, N_P p(1-p))$ and $B(N_H q, N_H q(1-q))$ respectively. At large population size, i.e., large $N_P$ and $N_H$, and when $p$, $q$ are not close to 0 and 1, these binomial distributions are reasonably approximated by Gaussian distributions. Since both $p$ and $q$ are not close to 0 and 1, we excluded the situation that probability $p$ and $q$ are both less than 0.1 in the same time. Another reason is that we consider that these links have less biological meaning than others. After that we obtain $N_{link}$ =758 links from the former 4005 links in the network. The sparsities (defined as the total number of edges in a network divided by the maximum number of possible edges) of ADHD and Control networks were 5.04% and 5.15% respectively.

Next, more specifically, the proportions of the patients/healthy controls in the population with a particular link, denoted by $\xi_P$ and $\xi_H$, follows Gaussian distributions $\xi_P \sim N(p, p(1-p)/N_p)$ and $\xi_H \sim N(q, q(1-q)/N_H)$ . Hence $\xi_P - \xi_H \sim N(p-q, p(1-p)/N_p + q(1-q)/N_H)$. Basically, $S = \xi_P - \xi_H$ is the score defined for this particular link. In real applications, the true values of $p$ and $q$ are never known. In our paper, the values of $p$ and $q$ are estimated by $\frac{L_P}{N_P}$ and $\frac{L_H}{N_H}$. It then comes the questions how large is *S* large enough to claim significance, and how variable the identified altered network could be at a pre-defined threshold for *S*. Under the null hypothesis, i.e., there is no difference between patients and healthy controls for a particular link, it is clear that $p = q$ and we can associate a *p*-value to the score *S* by $\Phi(-|\hat{S}|/\sigma)$, where $\Phi(\cdot) = \frac{1}{\sqrt{2\pi}}\int_{-\infty}^{\cdot} e^{-t^2/2}dt$ is the cumulative distribution function of the standard normal distribution, $\hat{S} = \hat{\xi}_P - \hat{\xi}_H$ is the estimated score from the data, and $\sigma^2 = p(1-p)/N_P + q(1-q)/N_H$. Conversely, a threshold for $|\hat{S}|$ to claim $\alpha$-level significance is given by $S_{th} = -\sigma \times \Phi^{-1}(\alpha)$.

Correction for multiple comparisons is needed to control false positive rate when a large number of links are tested simultaneously. We have a few methods to perform multi-comparison correction.

1. Bonferroni correction. When we correct for all the $N_{link} = 758$ links in the network using Bonferroni correction in 0.05-level significance, the score threshold will be $S_{th}^{Bonf} = -\sigma \times \Phi^{-1}(\frac{0.05}{N_{link}})$, which is much more stringent as shown in Fig S2. The saliency network is still statistically significant.
2. FDR. An alternative correction method for multiple comparisons is false discovery rate (FDR) procedure, which is less conservative than Bonferroni correction. In the analysis of this paper, the false discovery rate (FDR), is used to identify significant links. The score threshold will be $S_{th}^{FDR} = -\sigma \times \Phi^{-1}(P_{FDR})$, where $P_{FDR} = 5.85 \times 10^{-4}$ in $0.05$-level sig-nificance.

3. Permutation method: Nonparametric methods such as permutation can be used to assess the significance of the saliency circuit. The significance of the saliency network, after permutation test, is $p = 1.5 \times 10^{-5}$, is still statistically significant, after Bonferroni correction, at $0.05$-level.

Increasing population size will certainly increase the statistical power, as shown in Fig. S2. Once a score threshold is fixed, some links can be picked up as significantly altered and we can go into further follow-up analysis. Specifically, these links can form a network, marking the difference between patients and healthy controls more significant. Larger population size will help reduce this error probability.

A particular question will be the reproducibility of the network in a different dataset. Basically, for a particular link, one is concerned with the probability that $\hat{S}$ falls below the threshold $S_{th}$ when $|S| > S_{th}$, or $\hat{S}$ crosses the threshold $S_{th}$ when $|S| < S_{th}$. This can be estimated by the error probability $P_{error} = \Phi(-|S_{th} - \hat{S}| / \hat{\eta})$, where $\hat{\eta}^2 = \hat{\xi}_P(1-\hat{\xi}_P)/N_P + \hat{\xi}_H(1-\hat{\xi}_H)/N_H$. If this error probability is large, the resulting network is not stable and quite likely to be messed up when testing on a different dataset. If it is small for most of the links, the network should be reproducible. It is clear that this probability depends on both the population size and the distance from the score to the threshold. Links with scores close to the threshold have larger probability to vary in dif-ferent studies while those with score far from the threshold are more stable. In our network, the error probability $P_{error} < 0.05$ accounts for over 95.5 percent of links.

**Source Locations**

To examine whether each inter-voxel correlation differed significantly between healthy controls and ADHD patients, two-sample *t*-tests were performed for all inter-voxel correlations. Prior to the *t*-tests, a Fisher's r-to-z transformation was utilized to convert each correlation coefficient $r_{ij}$ into $z_{ij}$ to improve the normality. A significance level of $p$<0.05 was used to find out the voxels in insula and anterior cingular gyrus, between which functional connectivity was significant changed across the two groups. However, the functional connectivity of two voxels is very sensitive to noise; it can not work well to identify dysfunctional voxels only by thresholding a small significance level of *p* value. To solve this problem, for each voxel, a dysfunctional intensity is defined as following:

$$Intentisy = N^{A}_{p<0.05} / N^{B}$$

where $N^{A}_{p<0.05}$ is the number of voxels in region A that show significant different ($p$<0.05) functional connectivity with every voxel in region B compared with normal controls. $N^{B}$ is the total number of voxels in region B. This is reasonable since the value of intensity represents the significance of the changed correlation for each voxel. An intensity level (intensity>0.07) was further used to threshold voxels into two groups (unchanged part and changed part).

## DISCUSSION

The novel approach adopted here to identify altered functional circuits in the brains of ADHD patients has proved to be very informative. The approach is completely different from existing methods: seed-based analysis and independent component analysis and makes no assumptions about which circuits might be altered or that brain regions are independent of one another. Furthermore, our approach is similar to the GWAS approach, we have set a very strict test to assess the statistical significance of each found circuits. Our approach has identified the so called salience network, as the one showing the largest change in the whole dataset, although similar major changes also occurred in other three circuits. This involvement of the salience network has not, to the best of our knowledge, been found in previous studies. Interestingly, some of the main circuitry identified by other studies using an a priori seed-based approach, such as the links related to the putamen and caudate etc. (4, 5), was not found to be altered consistently in both patient groups. A link between the amygdala and hippocampus was present in BWAS approach, but absent after validations. Using a seed-based approach, Tian et al. (21) demonstrated that individuals with ADHD had increases in connectivity between the dorsal ACC and the frontoinsular cortex, the thalamus and the cerebellum. The dataset they used is actually a subset of ours. A subsequent study (22) used a 100-ROI brain parcellation and graph theoretical analysis to reveal altered small-world properties, suggestive of increased short-range connectivity and decreased long-range connectivity in the ADHD group. Another study reported a decrease in the correlation between the PCC and medial PFC, as well as a reduction in the negative correlation between the PCG and ACG (23), although our results do not find that they are significant here. Together, our findings from this study further support the idea that long-range disconnection might be a feature of ADHD (ORB to PCG and ITG to PHG).

Overall, our voxel-based morphometry analysis revealed no significant grey matter changes in any of the brain regions showing connectivity changes in ADHD patients. It therefore seems unlikely that observed changes were simply caused by reduced tissue volumes.

Although the current approach has only been applied to one of the major brain disorders, ADHD, it is clear that it could be easily applied to other forms of psychiatric, developmental or neurodegenerative disorders and provide information on how each of these disorders are characterized by a specific subset of functional connectivity changes as well as helping to identify possible common traits across, for example, affective or learning and memory disorders.

It could be argued that the changes in functional circuit we have iden-tified are simply a reflection of altered coherent activities (both positive and negative correlations) among brain regions in the resting state and that they might not be predictive of altered responsivity to internal or external stimuli promoting behavioral responses. For example, the salience network might enhance its coherency in the patients in the resting state, but return to its normal state and function normally in response to appropriate stimuli. This is certainly an issue for all resting state studies requiring further investigation although, as we will discuss below, there are some interesting parallels between our current findings and previous studies showing stimulus-evoked changes in these same circuits in ADHD patients.

So what might be the significance of the enhanced coupling we have found in the salience network of ADHD patients? This circuit enabes switching between the default mode and task-related state of brain connectivity, as recently proposed by Menon and Uddin (17, 18). In a network of six brain regions: anterior cingulate gyrus (ACG), anterior insula (AI), dorsolateral prefrontal cortex (DLPFC), posterior pariental cortex (PPC), ventromedial prefrontal cortex (VMPFC) and posterior cingulate cortex (PCC), ACG and AI are the saliency network, DLPRC and PPC act as the central-executive network, and VMPFC and PCC form the default mode network. They have argued that the salience network is to first identify stimuli from the vast and continuous stream of sensory stimuli that impact the sense. Once such a stimulus is detected, the anterior insula facilitates task-related information processing by initiating appropriate transient control signals to engage brain areas mediating attention, working memory, and higher code cognitive processes while disengaging the default mode network. The insula region is also reported to be involved in feelings of disgust as well as other emotions (24) and a recent fMRI study has shown enhanced responses in the insula to faces expressing disgust (25). A relationship between the different components of the salience network and various psychiatric and neurodegenerative diseases such as schizophrenia (26), Huntington's disease (27, 28) and depression (25) has already been reported.

The relationship between executive function and ADHD has long been speculated in the literature, but it lacks a neuroscience backup. There are a large numbers of articles and books refer to ADHD as a disorder of executive function of the mind (Brown, 2005). In particular, two conflicting views have emerged about how ADHD and executive function are related. One view holds that all individuals with ADHD suffer from significant impairment of executive function and ADHD is essentially a developmental impairment of executive function (29). The ADHD symptoms can be viewed as the evidence that the impairment of the central management networks that turn them on and off, but not with these fundamental cognitive functions themselves. The alternative view argued that some, but not all, suffer from significant impairments of executive function. These conflicting viewpoints come from divergent understandings of the nature of executive functions and how these functions should be assessed, and before the discovery of the salience network and its functions. Barkley's (1997) model was based not on data from neuropsychological tests of executive network, but on a conceptual framework derived primarily from integrating the crucial importance of language in human development. In Barkley model (1997) (30), he argued that ADHD is essentially impairment in the development of ability to inhibit. Interestingly, both views can be well explained and unified with our findings here. We have mentioned above that the salience network is essentially served as a dynamical switcher between two states: executive and default. They can be spontaneously activated and integrated by situational stimuli that, for the given individual, provide sufficient intrinsic satisfaction or threat to stimulate and sustain response. Due to the increasing link in the salience network, the network is easier to switch to the executive network, and hence harder to inhibit.

The saliency network may be involved in the control of other behaviors influenced by ADHD (see (18) for a review). Notably, individuals who have lesions on the insula can not update their prediction frameworks, but having the ability to judge the probability of events (31). The same brain regions also appear to be involved in prediction error and during risk evaluation predicted subsequent decision-making, as reported in (18) which implies that the insula plays a role in not only evaluating but also updating the probabilities of an outcome (32-35). The ACC is unlikely to be the sole region for cognitive control, but it plays a critical role in updating the prediction models and has been shown to be involved in both social and reward related associative learning (36-38).

The molecular and cellular basis of the saliency circuit is also well documented in the literature. Although it is believed that a large majority of ADHD cases arise from a combination of various genes, dopamine transporters seems play a central role. Candidate genes include dopamine transporter, dopamine receptors D2/D3, (39) dopamine beta-hydroxylase monoamine oxidase A, and the dopamine beta hydroxylase gene (DBH TaqI). (40) Dopamine has emerged as the primary neurochemical mediator in relation to various traits and behaviours mediated by the saliency network such as novelty-seeking, craving, nociception (41-44). Moreover, various studies have demonstrated the importance of dopaminergic modulation on the saliency network during executive tasks (45), suggesting that dopamine plays an important role in the function of the salience network. In (46, 47), it is found that the saliency network is the regions with a relatively high extrastriatal dopamine transporter. The saliency network function is directly affected by the synaptic availability of dopamine. A polymorphism that is shown to be associated with higher levels of dopamine transporters, which mediate dopamine transporters reuptake from the synaptic cleft into the presynaptic terminal, has been shown to be associated with greater activation of the insula and caudate along with deactivation of the cingulate during a verbal fluency task (48). A correlation between the binding of the D2/D3 ligand and grey matter density as measured by VBM has been observed in the anterior cingulate, insula and other regions (49), although we have not found the change of the grey matter across the salience network in individuals with ADHD significantly. Very recent PET experiments have also shown that both dopamine transporters and the D2/D3 receptor are significantly less abundant in the midbrain, hypothalamus, nucleus cacumbens, and caudate in ADHD patients, confirming that in dopamine reward signal is part of the causation of ADHD (50). A recent study (51) further suggests that it is not the dopamine transporter levels that indicate ADHD, but the brain's ability to produce dopamine itself. The study was done by injecting 20 ADHD patients and 25 controls with a radiotracer that attaches itself to dopamine transporters. The study found that it was not the transporter levels that indicated ADHD, but the dopamine itself. ADHD subjects showed lower levels of dopamine across the board. They speculated that since ADHD subjects had lower levels of dopamine to begin with, the number of transporters in the brain was not the telling factor. In summary, the insular dysfunction model of psychosis based on the salience network is con-sistent with the dopaminergic hypothesis of psychosis. Thus, the salience network provides a candidate cortical framework that is consistent with and builds on the existing dopaminergic hypothesis of ADHD. Despite the successful story as described above on dopaminegic relationships with the salience network, a GWAS approach to identify genes responsible for ADHD has proved less successful. Although twin and family studies have shown ADHD to be highly heritable, genetic variants influencing the trait at a genome-wide significant level has be carried out in (52). In a meta-study used data of sample size consisted of 2,064 trios, 896 cases and 2,455 controls, no genome-wide significant associations were found.

At cellular level, a special class of large bipolar spindle cells called Von Economo neurons (VEN) are common in the two regions of the salience network, different from the other brain regions. The dopamine D3 receptor is strongly expressed on the VEN, and it is natural to expect that these neurons involve in ADHD patients, as early suggested in Allman (2005) (29) and was supported by others. In a similar large population of ADHD patients to ours (218 ADHD, 358 controls), Shaw and his colleagues (53), using structural MRIs, have carefully compared the thickness of the left and right hemisphere of frono-insular area and surrounding cortex. In consistent with the estimated number of VENs in the right and left of frono-insular area and surrounding cortex in controls, Shaw et al. (53) found that the left, but not the right side was significantly thicker in ADHD, as in our findings here which indicates that the abnormality is mainly in the left side. Remember that these neurons are seen only in higher primates (54) and are thought to be involved in social processing in the wake of their rapid relaying ability (21, 23).

Another main pathway affected in both PU and NYU groups of ADHD patients was between the left and right TPO. This may reflect aspects of impaired temporal lobe ADHD can be very hard to live with. They can have gigantic mood swings, get very angry for almost no reason, and be nearly impossible to live with on a daily basis. The key to look for with this type of ADHD is anger outbursts for little or no reason. Decreased activity in the left temporal lobes causes problems with temper outbursts, aggressive behaviors, and even violence toward animals or other people.

The main focus of this study was to identify functional pathways altered in a large population of patients who were currently suffering from ADHD in order to try and help establish which ones are most strongly linked with ADHD per se. There were clearly a large number of differences between the patient groups which emphasizes the importance of not basing analyses of potential brain correlates of ADHD on a single type of patient group. In our study, we regress out the difference between gender difference, but we have not separated the gender groups. This is certainly an important factor that we have taken into account in our future study.

As reported before in a meta-analysis of relevant research based on 18 studies meeting inclusion criteria was performed (55), included primary symptomatology, intellectual and academic functioning, comorbid behavior problems, social behavior, and family variables. ADHD girls displayed greater intellectual impairment, lower levels of hyperactivity, and lower rates of other externalizing behaviors than ADHD boys. The need for future research examining gender differences in ADHD is strongly indicated, with attention to methodological limitations of the current literature, including the potential confounding effects of referral bias, comorbidity, developmental patterns, diagnostic procedures, and rater source.

## Acknowledgements

J. Feng is a Royal Society Wolfson Research Merit Award holder, partially supported by an EU grant BION, a UK EPSRC grant and National Centre for Mathematics and Interdisciplinary Sciences (NCMIS) in Chinese Academy of Sciences. We hereby appreciate what they, Carlton Chu, Virginia Tech's ARC, the ADHD-200 consortium and the Neuro Bureau (http://neurobureau.projects.nitrc.org/NeuroBureau/Welcome.html), have done for us.

## Conflict of interest

The authors declare no conflict of interest.

| Regions | **Abbr.** | **Regions** | **Abbr.** |
|---|---|---|---|
| Amygdala | AMYG | Orbitofrontal cortex (middle) | ORBmid |
| Angular gyrus | ANG | Orbitofrontal cortex (superior) | ORBsup |
| Anterior cingulate gyrus | ACG | Pallidum | PAL |
| Calcarine cortex | CAL | Paracentral lobule | PCL |
| Caudate | CAU | Parahippocampal gyrus | PHG |
| Cuneus | CUN | Postcentral gyrus | PoCG |
| Fusiform gyrus | FFG | Posterior cingulate gyrus | PCG |
| Heschl gyrus | HES | Precentral gyrus | PreCG |
| Hippocampus | HIP | Precuneus | PCUN |
| Inferior occipital gyrus | IOG | Putamen | PUT |
| Inferior frontal gyrus (opercula) | IFGoperc | Rectus gyrus | REC |
| Inferior frontal gyrus (triangular) | IFGtriang | Rolandic operculum | ROL |
| Inferior parietal lobule | IPL | Superior occipital gyrus | SOG |
| Inferior temporal gyrus | ITG | Superior frontal gyrus (dorsal) | SFGdor |
| Insula | INS | Superior frontal gyrus (medial) | SFGmed |
| Lingual gyrus | LING | Superior parietal gyrus | SPG |
| Middle cingulate gyrus | MCG | Superior temporal gyrus | STG |
| Middle occipital gyrus | MOG | Supplementary motor area | SMA |
| Middle frontal gyrus | MFG | Supramarginal gyrus | SMG |
| Middle temporal gyrus | MTG | Temporal pole (middle) | TPOmid |
| Olfactory | OLF | Temporal pole (superior) | TPOsup |
| Orbitofrontal cortex (inferior) | ORBinf | Thalamus | THA |
| Orbitofrontal cortex (medial) | ORBmed | | |

**Table 1**: The names and abbreviations of the regions of interest (ROIs).

| No. | Link Name | Score | $P$-value |
|---|---|---|---|
| 1 | INS.L-ACG.L | -0.1213 | $5.8856\times10^{-6}$ |
| 2 | ORBmid.L-IPL.L | 0.1246 | $1.2476\times10^{-4}$ |
| 3 | ORBmed.R-PCG.L | 0.094 | $2.2575\times10^{-4}$ |
| 4 | TPOsup.L-TPOsup.R | 0.1142 | $2.6881\times10^{-4}$ |
| 5 | REC.L-ANG.L | 0.0862 | $3.4125\times10^{-4}$ |
| 6 | MFG.L-ACG.R | -0.0851 | $3.8539\times10^{-4}$ |
| 7 | ORBsup.L-ORBmed.L | -0.1415 | $4.4194\times10^{-4}$ |
| 8 | ITG.R-PHG.R | -0.0769 | $4.513\times10^{-4}$ |
| 9 | AMYG.R-HIP.R | -0.1371 | $5.8545\times10^{-4}$ |

**Table 2**: Scores and $p$ value for 9 links after BWAS meta-analysis for the two data sets. Four links are highlighted which survive after the analysis procedure as described in Fig. 1.

| District | | Amount | Age/years Mean(SD ) | Gender | | IQ/scores Mean(SD ) | ADHD in-dex/scores Mean(SD ) |
|---|---|---|---|---|---|---|---|
| | | | | Females | Males | | |
| **PU** | Control | 143 | 11.43 (±1.86) | 59 | 84 | 118 (±13.15) | 29.34 (±6.41) |
| | ADHD | 102 | 12.09 (±2.04) | 10 | 92 | 106.03 (±13.10) | 50.38 (±8.39) |
| **NYU** | Control | 110 | 12.12 (±3.12) | 55 | 55 | 111 (±14.14) | 45.36 (±5.98) |
| | ADHD | 147 | 11.07 (±2.66) | 34 | 113 | 105.82 (±14.09) | 71.86 (±8.64) |
| **Meta-analysis** | Control | 253 | 11.73 (±2.50) | 114 | 139 | 115.06 (±13.98) | |
| | ADHD | 249 | 11.49 (±2.47) | 44 | 205 | 105.91 (±13.66) | |

**Table 3:** Data property of the two datasets used in our analysis

## Figure Legends

**Figure 1** The flow chart of BWAS analysis for ADHD datasets. A meta-analysis is performed for all data (from both Peking Univ. and NYU) and BWAS is applied to find the *p* value. In the end, there are 9 circuits which have a significant p value as listed in Tab. 2. On the other hand, two data sets are analyzed independently and there are 176 and 102 links are found with a *p* value smaller than 0.05. The common circuits among these three results are four common circuits as discussed in the text. Finally, source location algorithms are used to find the exact voxel location of each changed circuit, and the functional meaning of each circuit is analyzed.

**Figure 2** (A). A plot of six communities in different colours and altered links between patients and healthy controls, after BWAS analysis. Red lines: links increased in patients, blue lines, links present in normal controls. A total of nine links is identified. Line thickness corresponds to the score S. (B). A replot with their locations. Red lines: absolute value of the score *S* exceed 0.1, blue lines, absolute value of the scores *S* are less than 0.1

**Figure 3** Source locations of the saliency network. Warm colour part, altered voxels in either insula or anterior cingulate gyrus. Blue part is the unchanged part. It is clearly seen that the impaired part of insula is the anterior insula and anterior cingulate gyrus is the dorsal. The centre coordinates for the left insula and the left anterior cingulate gyrus source are denoted by green stars (the left side of the image corresponds to the right side of brain).

**Figure 4 (**A). In patients, significant positive partial correlation between intelligence tests scores and the partial correlation between TPOsup-R and TPOsup-L is found with r=0.2015 and p=0.0016. (B). In healthy control, however, there is no correlation between the intelligence tests scores and partial correlation between TPOsup-R and TPOsup-L with r=0.017 and p=0.7924.

**Fig. 1.**

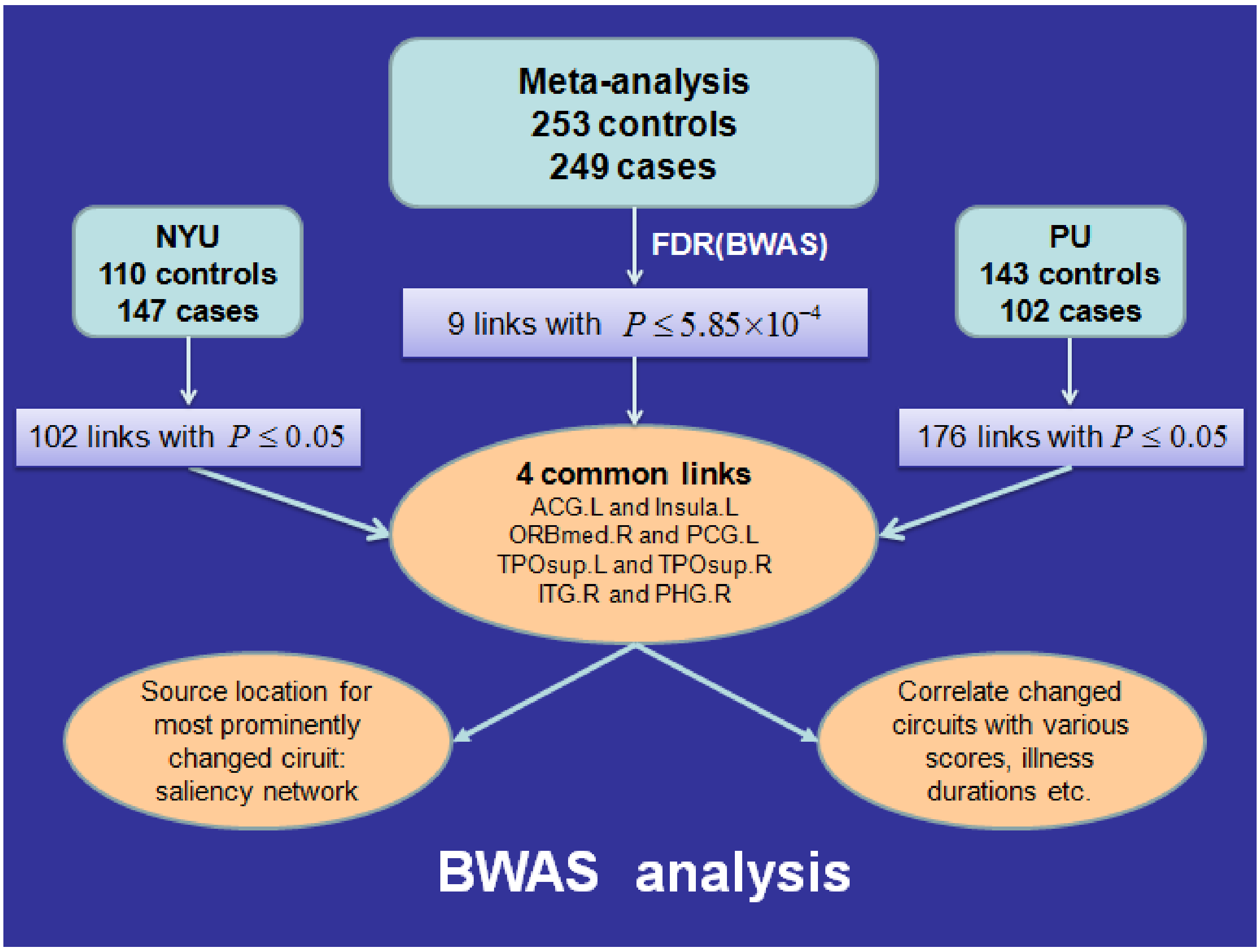
Meta-analysis
253 controls
249 cases
NYU
110 controls
147 cases
FDR(BWAS)
PU
143 controls
102 cases
9 links with $P \leq 5.85\times10^{-4}$
102 links with $P \leq 0.05$
176 links with $P \leq 0.05$
4 common links
ACG.L and Insula.L
ORBmed.R and PCG.L
TPOsup.L and TPOsup.R
ITG.R and PHG.R
Source location for most prominently changed ciruit: saliency network
Correlate changed circuits with various scores, illness durations etc.
BWAS analysis

**Fig. 2**

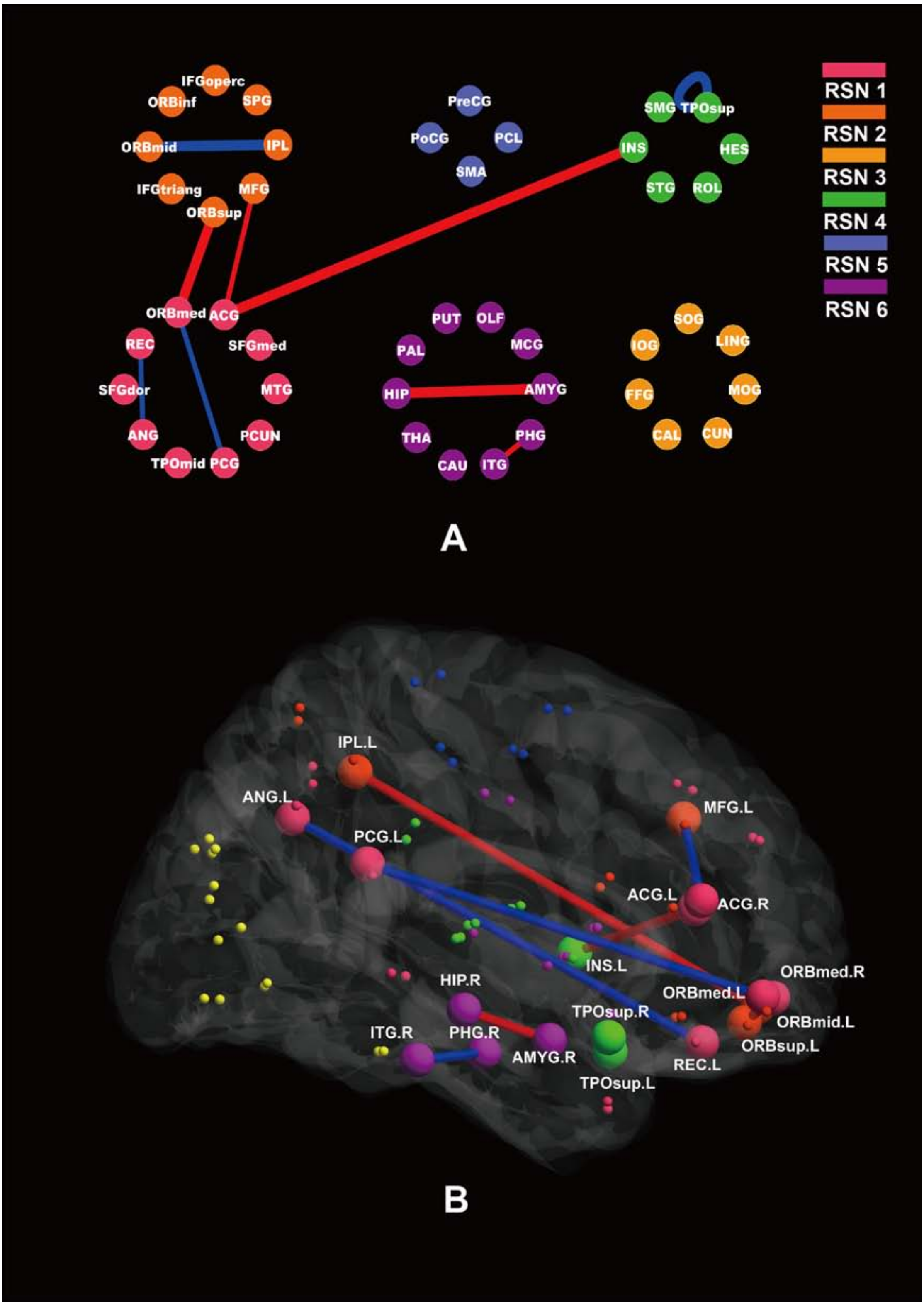

# Fig. 3

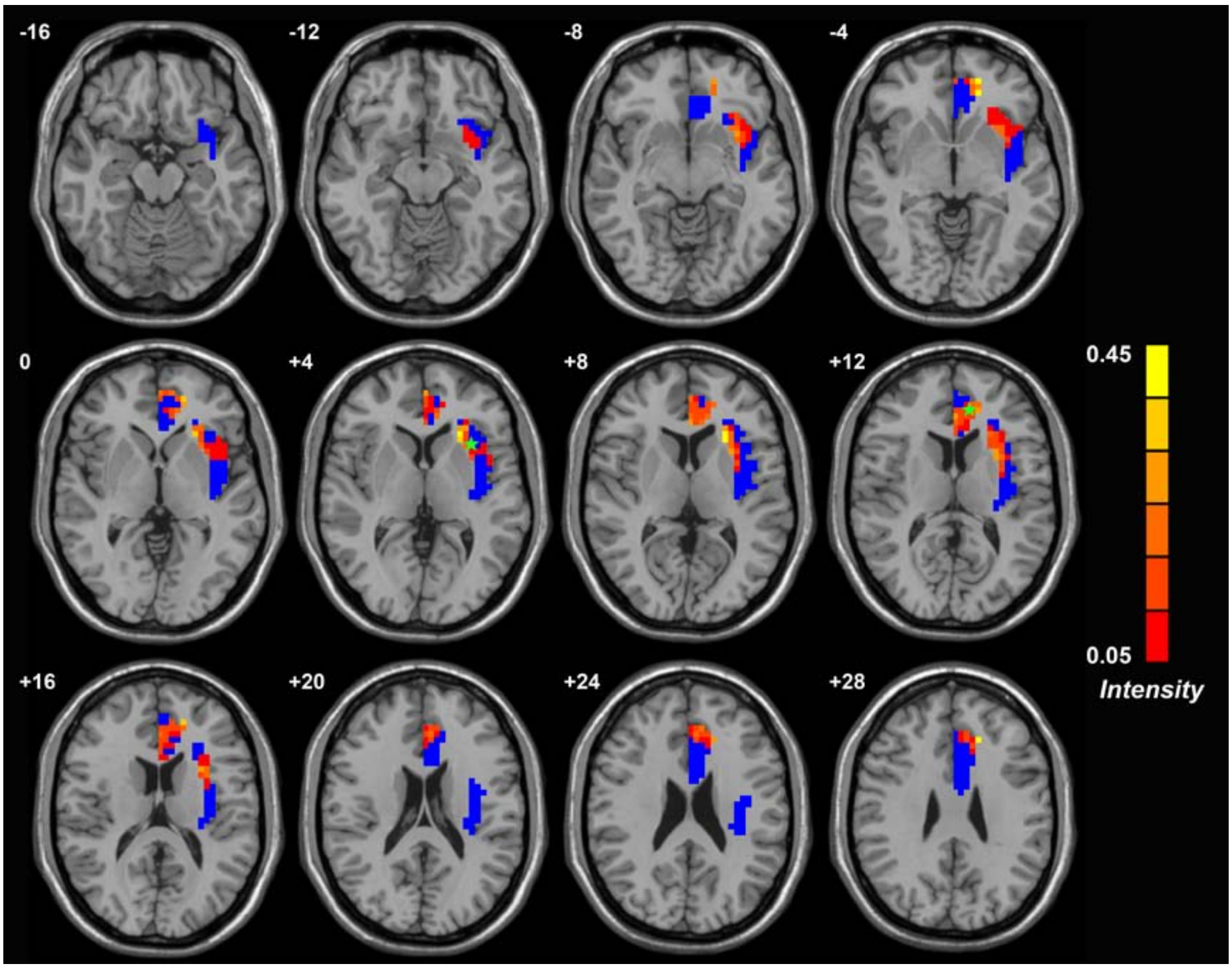

**Fig. 4**

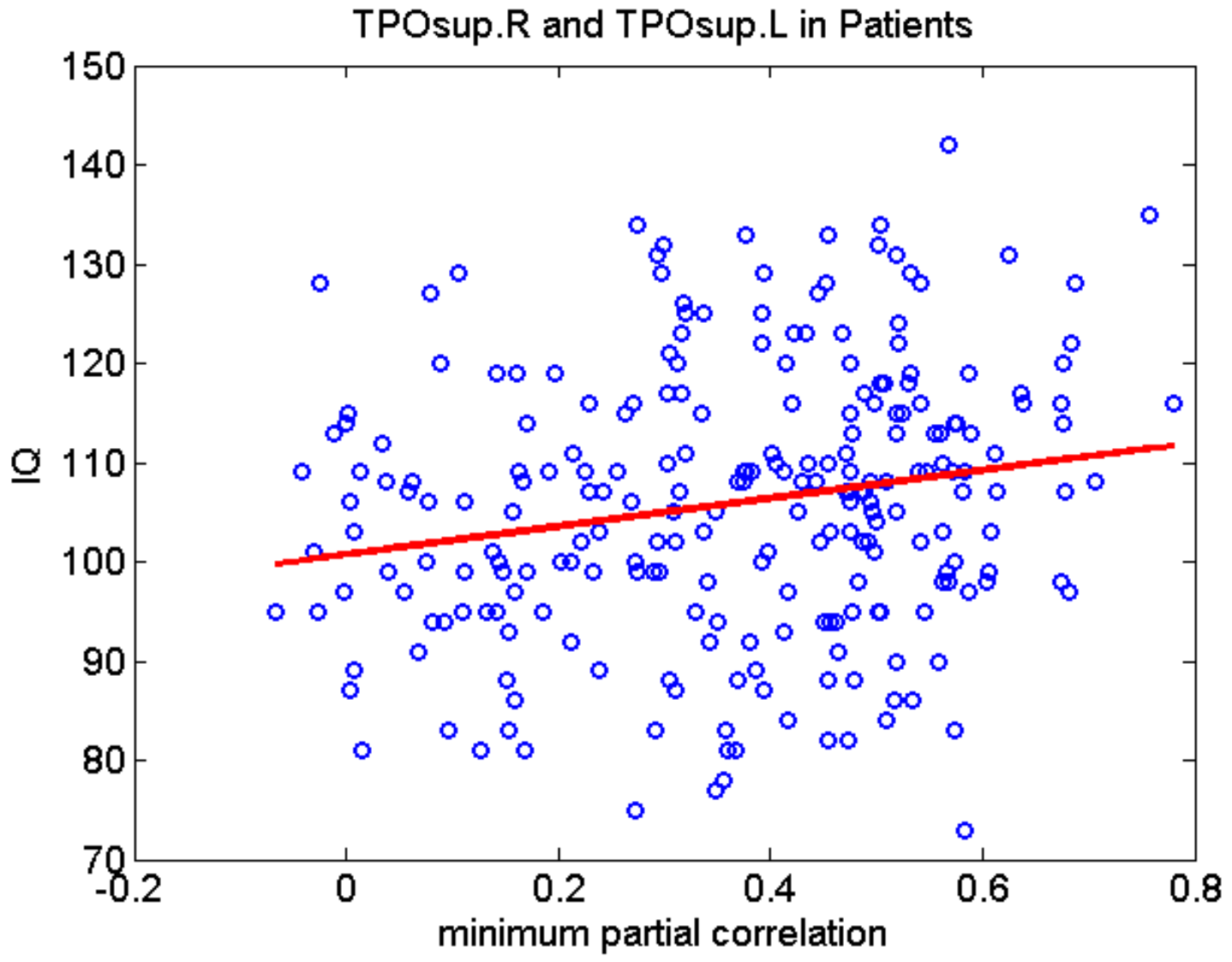

A

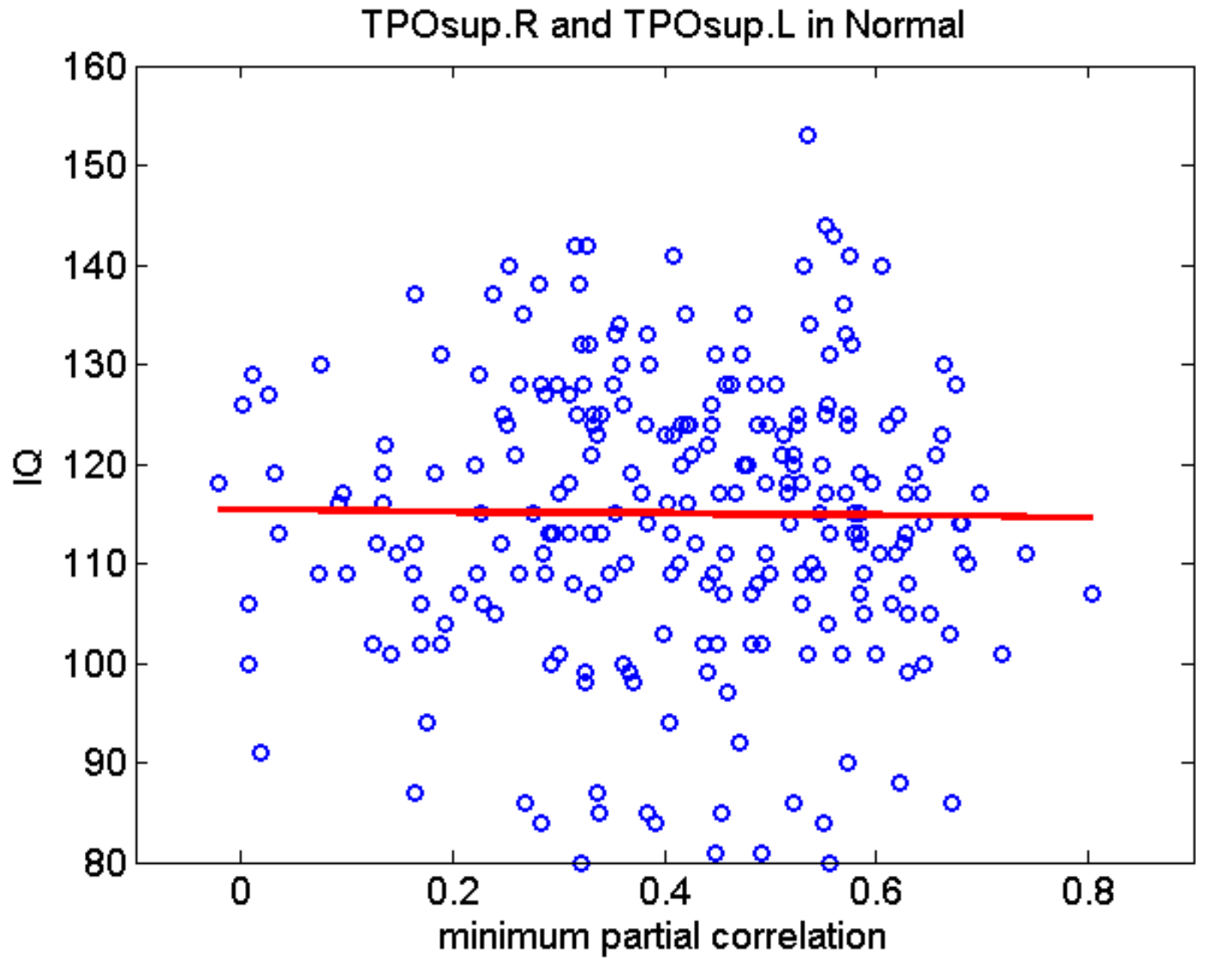

B

# Supplemental Materials:

## S1. Healthy template

A visualization of the six-community healthy controls and patients is shown in Fig. S1, where different colours correspond to the six communities as described in the main text (see Fig. 2A). The link between the corresponding brain regions of the left and right hemisphere is always there.

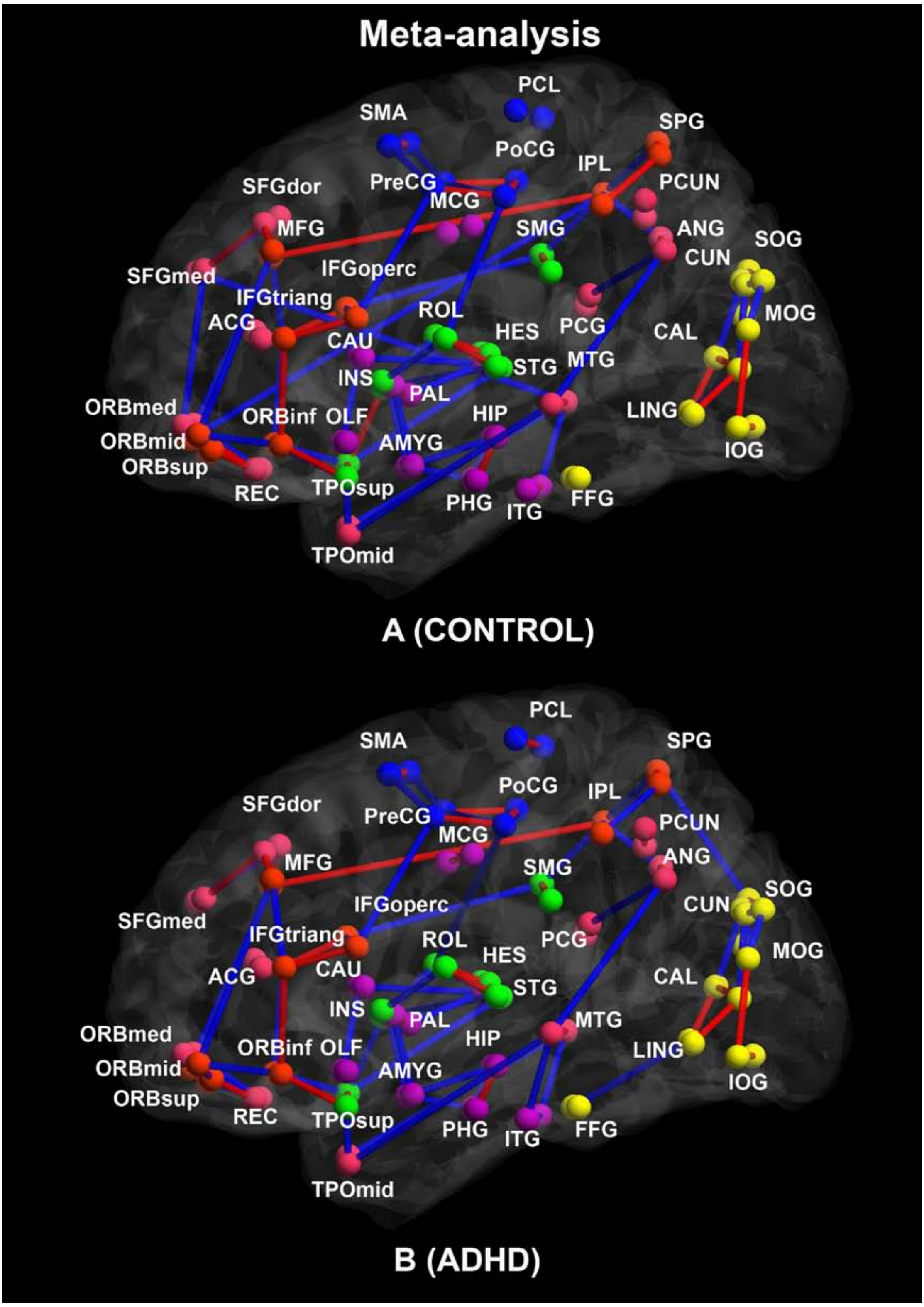

Fig. S1. (A) Six-community structure for control brain network of the 90 regions. (B) Corresponding structures in the ADHD brain.

## S2. Alternative BWAS meta-analysis:

Alternatively, we can carry out the analysis as in the flow-chat in Fig. S4 where we treat the NYU and PU group independently. As indicated, the saliency network survives as the only altered link.

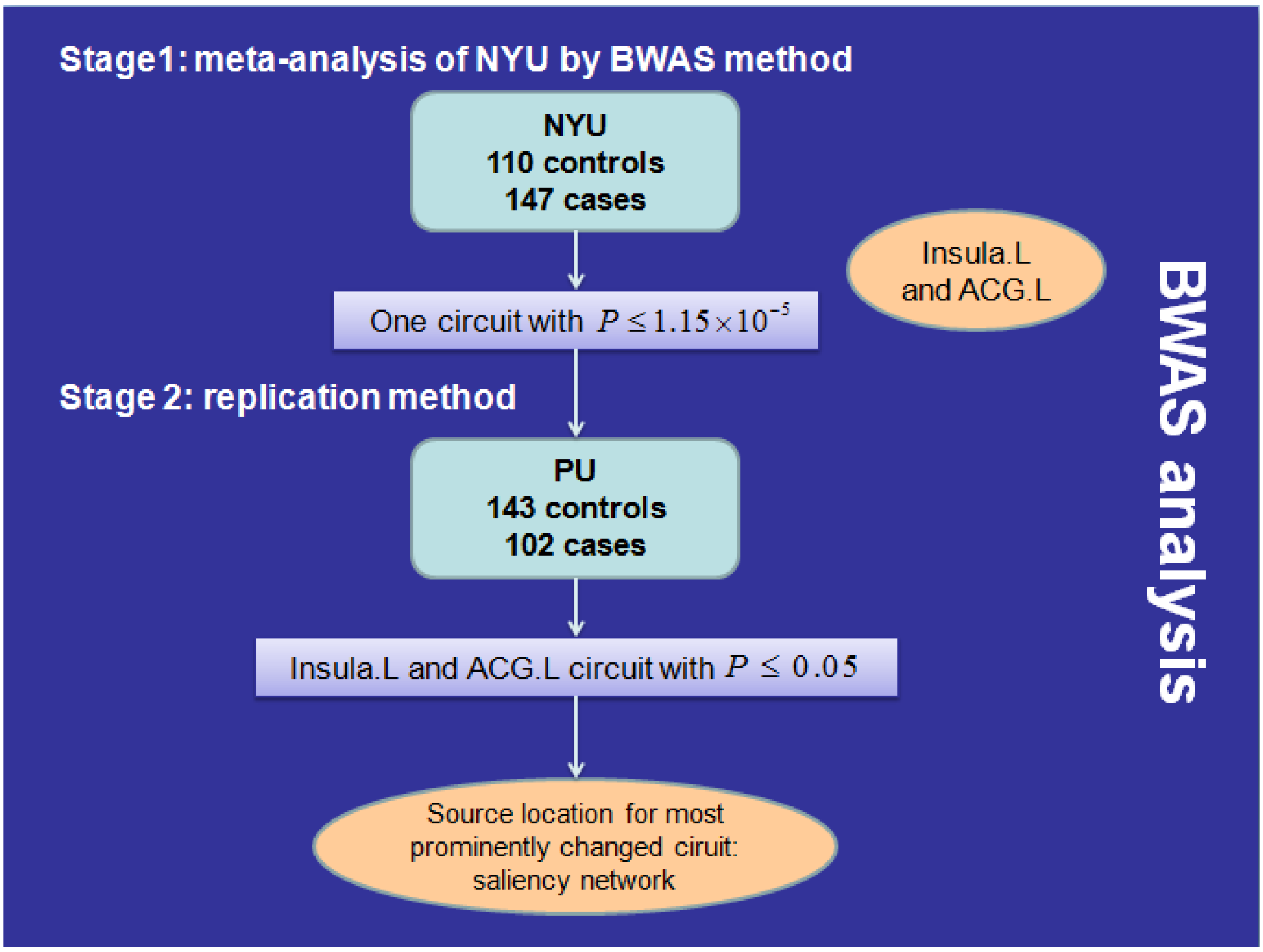

Fig. S2. An alternative way to perform BWAS. Note that here the replicate set (PU) is independent of the NYU set. The saliency network is still significant.

## S3. Table S1.

| Region | Difference | *P* value |
|---|---|---|
| Frontal_Inf_Tri_R | 0.0194 | 0.0016 |
| Frontal_Mid_R | 0.0205 | 0.0031 |
| Frontal_Mid_L | 0.0178 | 0.0056 |
| Hippocampus_L | 0.0153 | 0.0057 |
| Hippocampus_R | 0.0145 | 0.0074 |
| Postcentral_R | 0.0144 | 0.0116 |
| Temporal_Inf_R | 0.0216 | 0.0125 |
| Olfactory_R | 0.0221 | 0.0126 |
| Frontal_Inf_Tri_L | 0.0152 | 0.0146 |
| Frontal_Sup_R | 0.0136 | 0.0178 |

**T1 Data** The comparison of the mean of grey matter volumetric measures showed that the subjects in the ADHD group showed a 1.5% decrease in grey matter volume (*p* = 0.0953) in comparison with healthy controls. We did not detect any regions of increased grey matter in the ADHD group in comparison with the control group. Specifically, we found less grey matter volume in ADHD children in Inferior frontal gyrus, Middle frontal gyrus, Hippocampus, Postcentral gyrus, Inferior temporal gyrus, Olfactory cortex and Superior frontal gyrus. Ten most significant regions and corresponding *p*-values are listed in Table S1 (without correction).

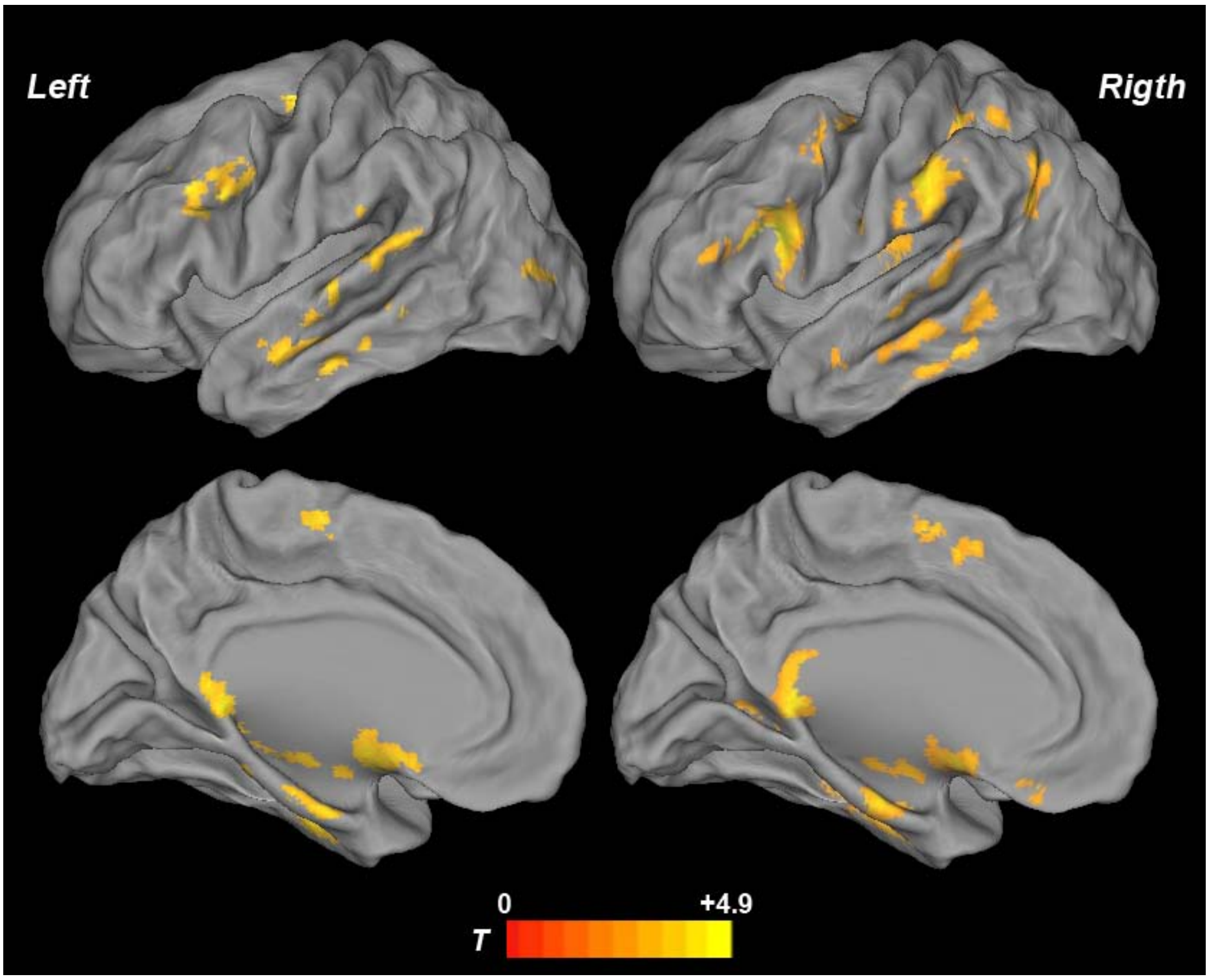

Fig. S3. The normalized, smoothed, segmented, modulated gray matter (GM) images were analyzed by using statistical parametric mapping (SPM8). Age and gender were included as nuisance variables for GM volume comparisons. Resultant t statistic maps were thresholded at $p < 0.05$ by using AlphaSim to correct for multiple comparisons. This figure displays the results of GM volume changes. Compared with healthy controls, ADHD patients had some regions with significantly reduced GM volume. Specifically, we found less GM volume in patients in inferior frontal gyrus, medial frontal gyrus, postcentral gyrus, middle temporal gyrus, hippocampus, paraHippocampal, fusiform gyrus. There was no region identified GM volume increases in patients compared with controls.

## S4. BWAS for the saliency network

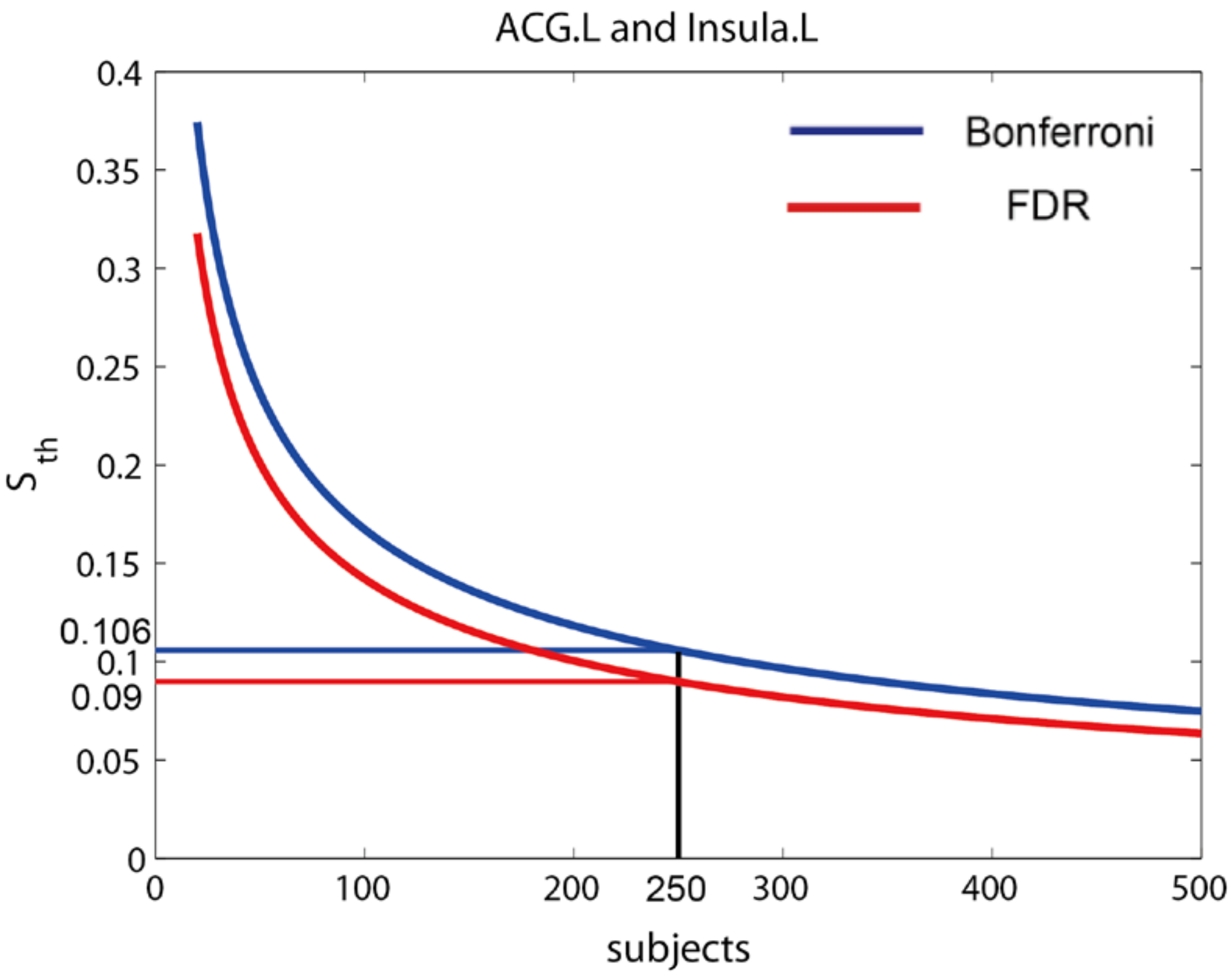

Figure S4: Bonferroni corrected and FDR corrected score threshold $S_{th}$ for the ACG.L and Insula.L circuit with different population size with $0.05$-level significance.